\documentclass[runningheads]{llncs}

\usepackage[T1]{fontenc}
\usepackage{graphicx}
\usepackage{amsmath,amssymb}
\usepackage{bm}
\usepackage{booktabs}
\usepackage{subcaption}
\usepackage{siunitx}
\usepackage{bbding}

\usepackage{marvosym}
\begin{document}
% -------------------------------------------------

\title{SQP-Based Cable-Tension Allocation for Multi-Drone Load Transport}

\author{Lamberto Vazquez-Soqui\inst{1} \and
Fatima Oliva-Palomo\inst{1}\textsuperscript{(\Letter)} \and
Diego Mercado-Ravell\inst{2} \and
Pedro Castillo\inst{3}
}
\authorrunning{L. Vazquez-Soqui et al.}

\institute{Center for Research in Mathematics (CIMAT AC), campus Zacatecas, {Calle Lasec y Andador Galileo Galilei, Manzana 3, Lote 7 Quantum Ciudad del Conocimiento}, 
 {Zacatecas},
 {98160}, 
 {Zacatecas},
 {Mexico}, \email{lamberto.vazquez@cimat.mx}, \email{fatima.oliva@cimat.mx} \and Center for Research and Advanced Studies (CINVESTAV-IPN), campus Guadalajara, { Av. del Bosque 1145, El Bajío, 45017 Zapopan, Jalisco}, 
 {Mexico}, \email{diego.mercado@cinvestav.mx}
 \and
Université de Technologie de Compiègne, CNRS, Heudiasyc UMR 7253, Compiègne, 60200, France, \email{pedro.castillo@hds.utc.fr}
 }

\maketitle

% -------------------------------------------------
\begin{abstract}
Multi-Agent Aerial Load Transport Systems (MAATS) offer greater payload capacity and fault tolerance than single-drone solutions. However, they have an underdetermined tension allocation problem that leads to uneven energy distribution, cable slack, or collisions between drones and cables. This paper presents a real-time optimization layer that improves a hierarchical load-position-attitude controller by incorporating a Sequential Quadratic Programming (SQP) algorithm. The SQP formulation minimizes the sum of squared cable tensions while imposing a cable-alignment penalty that discourages small inter-cable angles, thereby preventing tether convergence without altering the reference trajectory. We tested the method under nominal conditions by running numerical simulations of four quadrotors. Computational experiments based on numerical simulations demonstrate that the SQP routine runs in a few milliseconds on standard hardware, indicating feasibility for real-time use. A sensitivity analysis confirms that the gain of the cable-alignment penalty can be tuned online, enabling a controllable trade-off between safety margin and energy consumption with no measurable degradation of tracking performance in simulation. This framework provides a scalable path to safe and energy-balanced cooperative load transport in practical deployments.

\end{abstract}

\section{Introduction}\label{sec:intro}
The use of Unmanned Aerial Vehicles (UAVs) for cooperative payload transportation has matured from early architectural and formation-based strategies~\cite{Maza2010} to a viable alternative with implementations in multiple domains~\cite{Jackson2020,Mercado2024,oliva2024}. By distributing the lift task among several low-cost platforms, these Multi-Agent Aerial Load Transportation Systems (MAATS) overcome the thrust-to-weight limits of single quadrotor solutions. They also enhance fault tolerance; if one vehicle fails, the others can maintain the load airborne long enough to execute a controlled landing, significantly reducing the risk of damage. The development of these systems has been accelerated even more thanks to new simulation tools~\cite{Li2023} and cooperative control methods~\cite{Geng2020,Villa2021}.

Recent surveys have identified three persistent technical challenges that impede routine deployment~\cite{VillaBrandao2020}: first, the force-balance equations are underdetermined, so the controller must select tensions carefully to avoid energy imbalance, cable slack, or unsafe configurations; second, the payload behaves as an underdamped pendulum, amplifying control effort and uneven battery discharge; and third, with more than two or three UAVs, the risk of inter-vehicle or cable collisions rises sharply. Different research groups have investigated these problems using nonlinear control strategies to improve payload stability~\cite{Goodarzi2015}, robust hierarchical geometric controllers~\cite{Cai2024}, and new frameworks for aerial manipulation.

Several optimization-centered solutions incorporate fast numerical solvers into the system. Wahba and Hönig proposed a force allocation strategy that solves a cascade of small quadratic programs to compute cable tensions, achieving a desired payload wrench~\cite{Wahba2024}. Wehbeh \textit{et al.} introduced a fully distributed model-predictive controller in which each quadrotor solves its own MPC locally and shares only a compact state vector with its teammates at every step, achieving stable four-UAV cooperative payload transport over a low-bandwidth radio link \cite{Wehbeh2020}. Robustness has also been addressed through adaptive formation-based control strategies: Doakhan \emph{et al.} proposed a hierarchical controller that combines finite-time sliding mode and adaptive laws to maintain a quadrotor formation while ensuring load transport stability under unknown payload mass and external disturbances~\cite{Doakhan2023}. Jackson \emph{et al.} presented a scalable distributed trajectory optimization approach that parallelizes computation over multiple computationally-limited agents while handling nonlinear dynamics and non-convex constraints~\cite{Jackson2020}. Cooperative transport has also been explored through fast trajectory optimization techniques for agile cable-suspended payload maneuvers~\cite{Foehn2016} and decentralized co-manipulation strategies based on internal force optimization~\cite{Tognon2018}. These works highlight the importance of explicit optimization in addressing redundancy, pendulum swing, and safety, rather than relying solely on tuned gains.

Despite these advancements, several gaps remain, particularly in how current tension allocators address the simultaneous balancing of force and cable-to-cable alignment, as most focus only on minimizing an energy surrogate and enforcing geometric separation or specifying a formation. On the other hand, more complex cost functions impose prohibitive computation burdens. The present work aims to fill those gaps by implementing a Sequential Quadratic Program (SQP) as an optimization layer that augments the hierarchical control structure presented by Oliva-Palomo~\cite{oliva2024}, explicitly addressing both tension balancing and cable-to-cable alignment without restricting the vehicles' operating envelope.

The remainder of this paper is organized in this manner. The hierarchical control architecture is shown in Section \ref{sec:architecture}. Section \ref{sec:optimization} introduces the main contribution: an SQP-based tension allocation formulation that distributes force by minimizing squared cable tensions, while incorporating a cable alignment penalty that discourages drones from getting too close to each other. This formulation utilizes quadratic subproblems that admit closed-form solutions, enabling real-time execution. Section \ref{sec:results} describes the simulation setup and validates the approach through numerical simulations of four quadrotors in MATLAB/Simulink, demonstrating trajectory tracking performance, balanced tension distribution, safety improvements through maintained cable angles, and computational feasibility averaging less than \SI{2}{\milli\second} per optimization cycle. Section \ref{sec:conclusion} draws some conclusions.

\section{Hierarchical Control Architecture}
\label{sec:architecture}

\begin{figure}[t]
 \centering
 \includegraphics[width=0.75\linewidth]{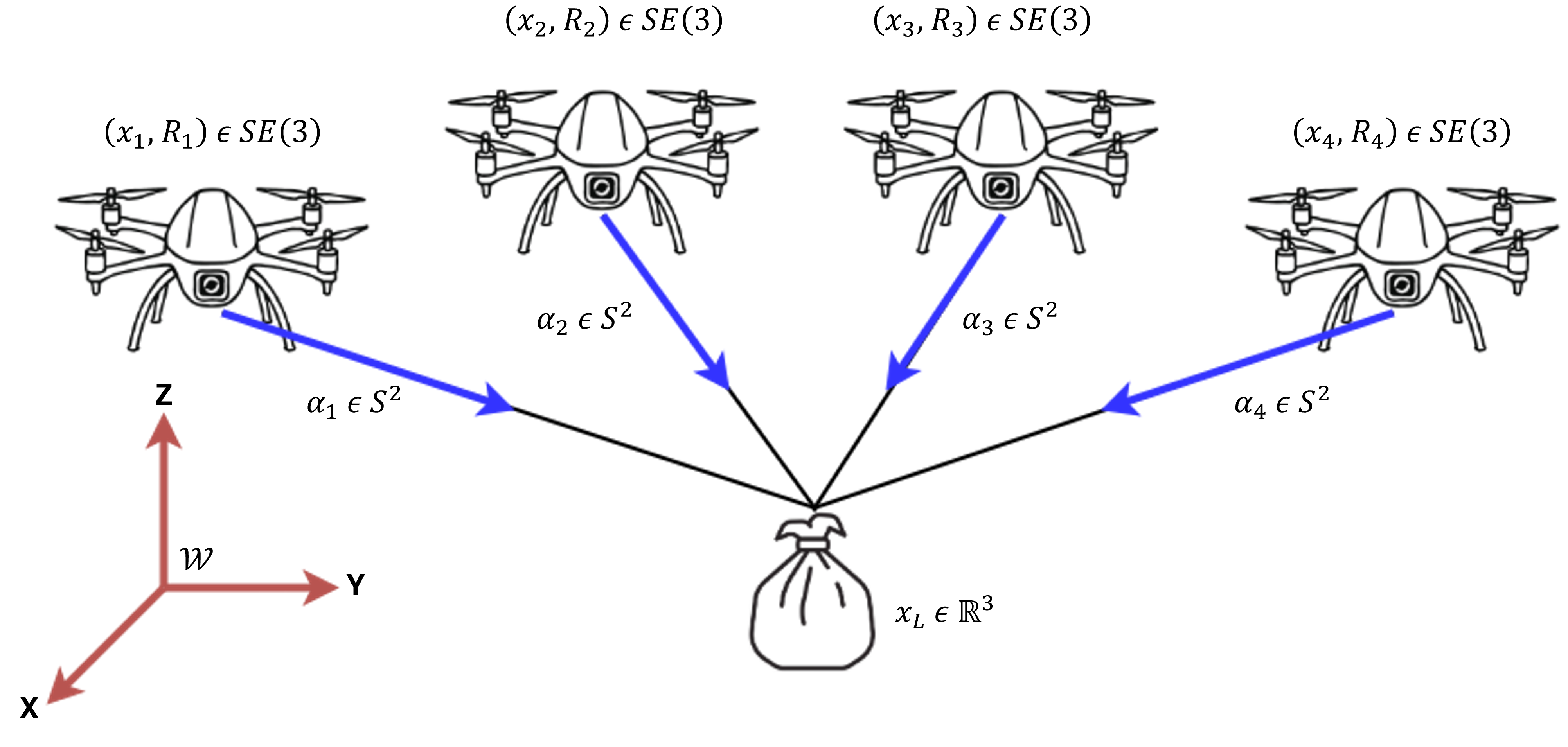}
 \caption{Multi-Agent Aerial Transportation System (MAATS). Nomenclature and reference frames for an n-UAV MAATS.}
 \label{fig:nomenclature}
\end{figure}

\begin{figure}[t]
 \centering
 \includegraphics[width=0.67\linewidth]{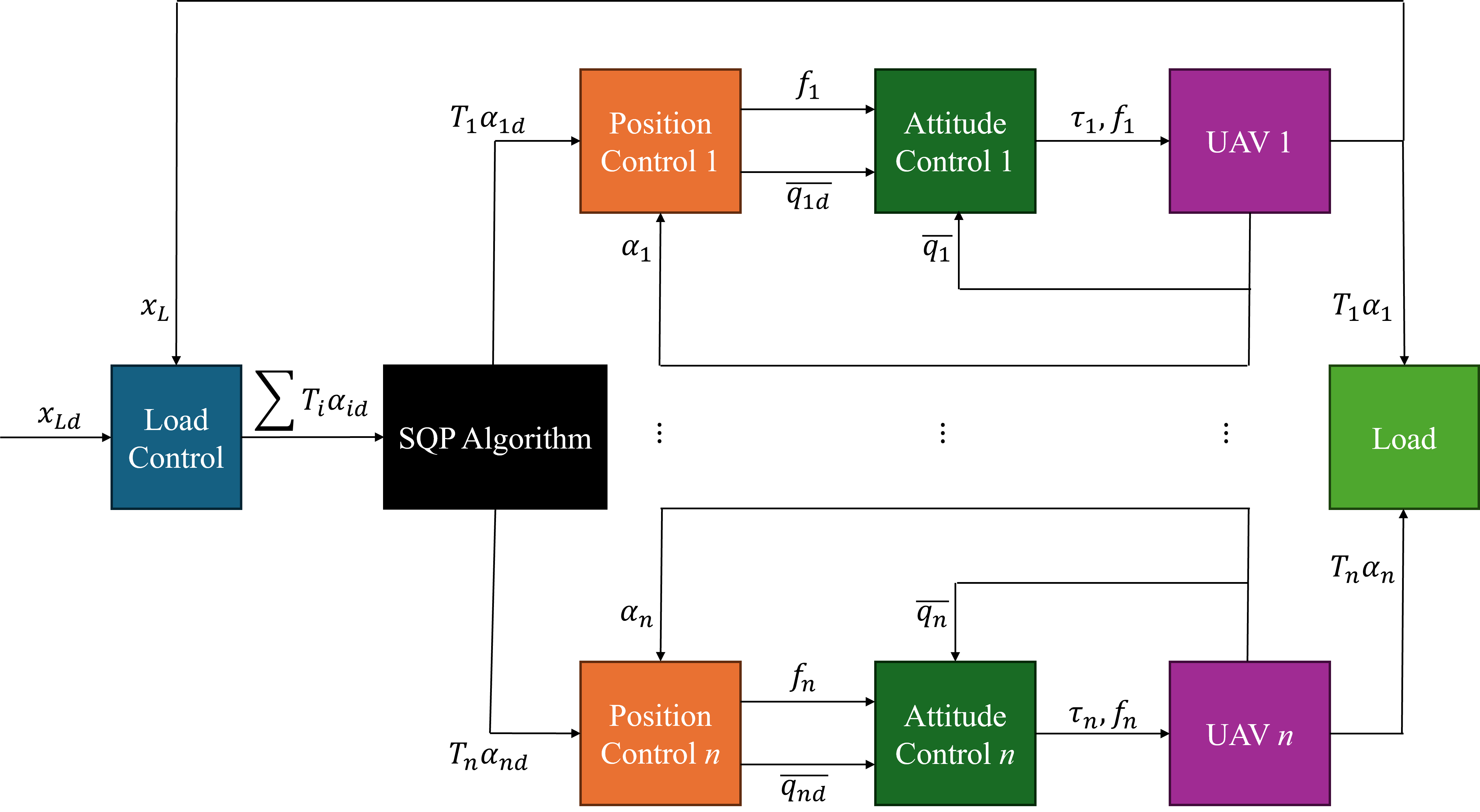}
 \caption{Block Diagram of the Control Strategy in a four-layer cascade.}
 \label{fig:block}
\end{figure}

We adopt the four-layer cascade controller of Oliva-Palomo \textit{et al.} \cite{oliva2024}. Figure \ref{fig:nomenclature} collects the notation and frames, while Fig. \ref{fig:block} shows the corresponding block diagram. The hierarchical architecture consists of load control that generates the total virtual force for trajectory tracking, tension allocation that distributes this force among UAVs, position control for each vehicle, and attitude control that generates final thrust and torque commands. The SQP algorithm (black box) serves as the tension allocation layer, computing individual tension magnitudes $T_{id}$ and directions $\alpha_{id}$ for each UAV from the virtual controller $\sum T_{id}i \alpha_{id}$. The feedback structure ensures that the actual tensions of the cable stay within the safety limits while following these optimized references.

\subsection{System Dynamics}

Consider $n$ UAVs that transport a point mass through rigid, massless cables of length $L_i$. Each UAV $i\in\{1,\dots,n\}$ has mass $m_i$ and inertia matrix $J_i$. The translational model for each UAV is
\begin{equation}
 m_i\ddot{x}_i=f_i R_i(\bar{q}_i)e_3-m_i g e_3+T_i\alpha_i,
\end{equation}
where $x_i\!\in\!\mathbb{R}^3$ is the position of the UAV $i$ in the inertial frame $\mathcal{W}$, $f_i$ is the total thrust magnitude, $R_i(\bar{q}_i)\!\in\!SO(3)$ is the rotation matrix parameterized by the unit quaternion $\bar{q}_i=[\eta_i,\epsilon_i^\top]^\top$ with scalar part $\eta_i$ and vector part $\epsilon_i$. The vector $e_3=[0,0,1]^\top$ points along the $z$-axis of $\mathcal{W}$, $g$ is gravitational acceleration, $T_i$ is the tension magnitude in the cable $i$, and $\alpha_i\!\in\!S^2$ is the unit vector pointing from UAV $i$ to the load. The rotational dynamics are
\begin{equation}
 \dot{\bar{q}}_i=\tfrac{1}{2}\bar{q}_i\otimes\bar{\Omega}_i,
\end{equation}
\begin{equation}
 J_i\dot{\Omega}_i+\Omega_i\times J_i\Omega_i=\tau_i,
\end{equation}
with body angular velocity $\Omega_i\!\in\!\mathbb{R}^3$, pure quaternion $\bar{\Omega}_i=[0,\Omega_i^\top]^\top$, control torque $\tau_i\!\in\!\mathbb{R}^3$, quaternion multiplication $\otimes$ as used in \cite{oliva2024}, and cross product $\times$.

The rigid cable constraint relates the UAV and load positions:
\begin{equation}
 x_i=x_L-L_i\alpha_i, \quad \lVert\alpha_i\rVert=1
\end{equation}

where $x_L\!\in\!\mathbb{R}^3$ is the payload position. The load (mass $m_L$) dynamics are
\begin{equation}
 m_L\ddot{x}_L=-m_L g e_3-\sum_{i=1}^{n}T_i\alpha_i.
\end{equation}

\subsection{Load Controller}
The load-level controller guarantees trajectory tracking for the suspended mass. With tracking error $e_L=x_L-x_{Ld}$ and its derivative $\dot{e}_L=\dot{x}_L-\dot{x}_{Ld}$, where $x_{Ld}$ is the desired payload trajectory,
\begin{equation}
 u_L=-m_L(g e_3+\ddot{x}_{Ld})-k_{pL}e_L-k_{dL}\dot{e}_L-k_{iL}\!\!\int e_L \, dt,
\end{equation}
where $k_{pL},k_{dL},k_{iL}\!\in\!\mathbb{R}^{3\times3}$ are diagonal positive-definite gain matrices. This virtual control must equal the negative sum of all desired cable tensions:
\begin{equation}
u_L = -\sum_{i=1}^{n}T_{id}\alpha_{id},
\label{eq:load_control}
\end{equation}
where $T_{id}$ and $\alpha_{id}$ are the desired tension magnitude and direction of cable $i$.

\subsection{Position Controller}

Let $x_{ei}=x_i-x_{id}$ be the position error of UAV $i$ relative to its desired position $x_{id}=x_{Ld}-L_i\alpha_{id}$. The virtual force applied to UAV $i$ is  
\begin{equation}
 u_i=m_i(g e_3+\ddot{x}_{Ld})-T_{id}\alpha_{id}
    -k_{pi}x_{ei}-k_{di}\dot{x}_{ei}-k_{ii}\!\!\int x_{ei}\,dt,
\end{equation}
where $(k_{pi},k_{di},k_{ii})\!\in\!\mathbb{R}^{3\times3}$ are diagonal positive-definite gain matrices.

The desired thrust magnitude and direction are extracted as
\begin{equation}
f_{id}=\lVert u_i\rVert, \quad \hat{u}_{id}=u_i/\lVert u_i\rVert.
\end{equation}

Assuming zero desired yaw, the reference quaternion becomes  
\begin{equation}
 \bar{q}_{id}=\begin{bmatrix}
  \tfrac{1}{2}\sqrt{2\hat{u}_{id,3}+2}\\
  -\hat{u}_{id,2}/\sqrt{2\hat{u}_{id,3}+2}\\
   \hat{u}_{id,1}/\sqrt{2\hat{u}_{id,3}+2}\\
  0
 \end{bmatrix},
\end{equation}
where $\hat{u}_{id,j}$ denotes the $j$-th component of $\hat{u}_{id}$. The desired angular velocity $\Omega_{id}$ is obtained by differentiating $\hat{u}_{id}$.

\subsection{Attitude Controller}
The desired attitude $\bar{q}_{id}$ and body rate $\Omega_{id}$ of the upper layers define the quaternion error and the angular velocity error
\begin{equation}
 \bar{q}_{ie}=\bar{q}_{id}^*\otimes\bar{q}_i, \quad \Omega_{ie}=\Omega_i-\Omega_{id}.
 %=\begin{bmatrix}q_{id0}q_{i0}+q_{id}^{\!T}q_i\\
 %         q_{id0}q_i-q_{i0}q_{id}-q_{id}\!\times\!q_i
 % \end{bmatrix},
\end{equation}
where $\bar{q}_{id}^*=[q_{id0},-q_{id}^T]^T$ is the conjugate. %and the angular velocity error 
%\begin{equation}
%\Omega_{ie}=\Omega_i-\Omega_{id}.
%\end{equation}
With error manifold
\begin{equation}
 s_i=\Omega_{ie}+\rho_i q_{ie},
\end{equation}
where $\rho_i\!\in\!\mathbb{R}^{3\times3}$ is a positive definite diagonal and $q_{ie}$ is the vector part of $\bar{q}_{ie}$, the torque command is
\begin{equation}
 \tau_i=-K_{di}s_i-\beta_i\operatorname{sat}(\gamma_i s_i),
\end{equation}
where $K_{di},\beta_i,\gamma_i\!\in\!\mathbb{R}^{3\times3}$ are positive-definite gains and $\operatorname{sat}(\cdot)$ is component-wise saturation.

The hierarchical architecture above provides the control framework for multi-UAV transportation; however, the critical allocation problem remains: how to distribute the total control effort $u_L$ among the $n$ agents. Equation \eqref{eq:load_control} represents an underdetermined system with equations $3$ and unknowns $3n$, which admit infinitely many solutions. The following section addresses the tension allocation problem through an SQP formulation that leverages this redundancy to optimize system performance while respecting physical constraints.

\section{SQP-Based Tension Allocation}
\label{sec:optimization}

Given the virtual control input $u_L\in\mathbb{R}^3$ from the load controller, we seek the desired cable tensions $T_{id}\ge 0$ and unit directions $\alpha_{id}\in S^2$ that satisfy
\begin{subequations}
\begin{align}
\min_{T_{id},\,\alpha_{id}}\;& 
      J \;=\; \tfrac12\sum_{i=1}^{n} T_{id}^{2} 
            + \mu\!\!\sum_{i=1}^{n-1}\sum_{j=i+1}^{n}(\alpha_{id}^{\!\top}\alpha_{jd})^{2}
            \label{eq:obj}\\[2pt]
\text{s.t. }\;& 
      \sum_{i=1}^{n} T_{id}\,\alpha_{id} \;=\; -u_L                       \label{eq:force_balance}\\
&     T_{id} \ge 0                                                      \label{eq:tension_bounds}\\
&     \|\alpha_{id}\|^{2}=1                                             \label{eq:unit_norm}
\end{align}
\label{eq:optim_problem}%
\end{subequations}
Here, \(i\) indexes the \(n\) cables, while \(j\) denotes the other cables in each pairwise term.

This formulation treats both tensions and directions as decision variables, allowing the optimizer to trade off energy use against safety spacing between agents.

The first term in \eqref{eq:obj} minimizes the sum of squared tensions, promoting both minimal total effort and balanced distribution among all agents. However, using only this quadratic term leads to solutions where cables converge in similar directions, increasing the risk of collisions and reducing controllability.

The second term, weighted by $\mu$, addresses these concerns through a cable alignment penalty. The dot product $\alpha_i^\top \alpha_j$ equals $\cos(\theta_{ij})$, where $\theta_{ij}$ is the angle between cables $i$ and $j$. When $\theta_{ij} \to 0$, this term approaches one, penalizing such configurations. The quadratic form ensures that the optimization space is smooth and differentiable while providing symmetric treatment of parallel and antiparallel configurations.

\subsection{Sequential Quadratic Programming}
\begin{figure}[t]
 \centering
 \includegraphics[width=0.725\linewidth]{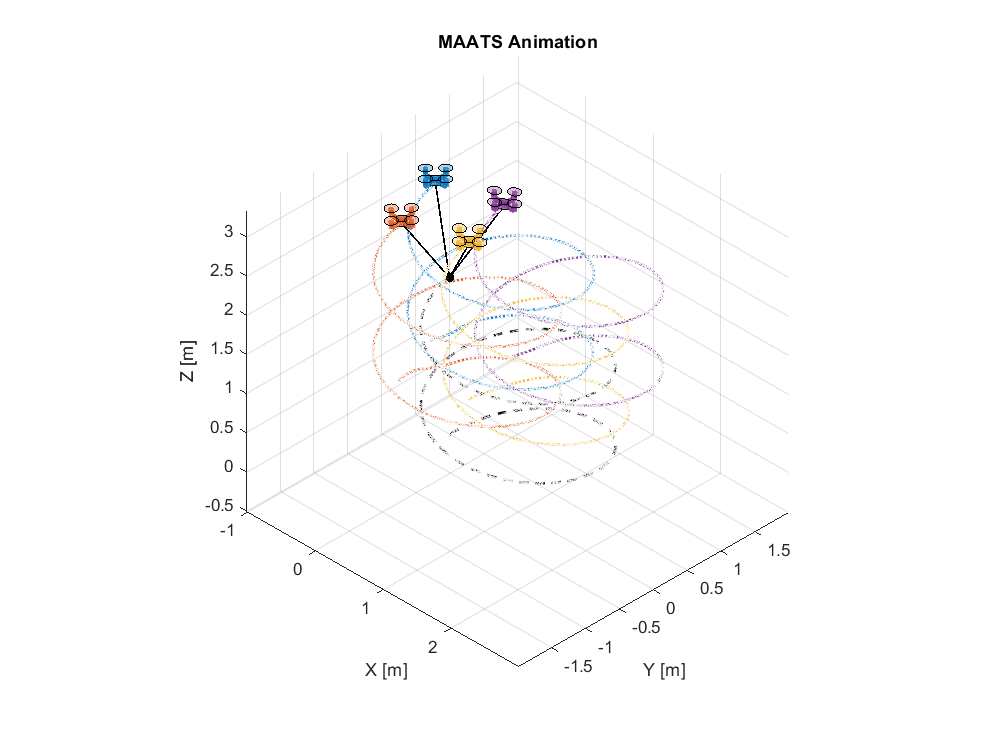}
 \caption{Three-dimensional view of the four-UAV system transporting the payload during the spiral trajectory. The SQP allocator maintains safe cable angles while ensuring balanced tension distribution.}
 \label{fig:3d_config}
\end{figure}
Sequential Quadratic Programming (SQP) solves the nonlinear program~\eqref{eq:optim_problem} by iteratively building a quadratic approximation of the Lagrangian and a first-order linearization of the constraints \cite{MatlabSQP}.  Each iteration therefore, yields a quadratic program that can be solved efficiently, even with the bilinear force balance~\eqref{eq:force_balance} and the unit norm constraints~\eqref{eq:unit_norm}.

A critical feature of our implementation is the warm-start initialization strategy. This approach ensures continuity between control cycles, preventing jumps in cable configurations that could cause payload oscillations. Furthermore, starting near the previous optimal solution dramatically reduces the number of iterations required for convergence, which is crucial for real-time performance. Initialization also maintains physical consistency, allowing UAVs to retain their relative positions unless force requirements demand reconfiguration.

The objective gradient with respect to our stacked decision vector $z = [T_1, ..., T_n, \alpha_1^T, ..., \alpha_n^T]^T$ exhibits a specific structure that the algorithm exploits:
\begin{equation}
\nabla J(z) = \begin{bmatrix}
T_1 \\
\vdots \\
T_n \\
2\mu\sum_{j \neq 1}(\alpha_1^T \alpha_j)\alpha_j \\
\vdots \\
2\mu\sum_{j \neq n}(\alpha_n^T \alpha_j)\alpha_j
\end{bmatrix}
\label{eq:gradient}
\end{equation}

The first $n$ components of \eqref{eq:gradient}, corresponding to the derivatives with respect to tensions, are simply the tensions themselves from the quadratic term in \eqref{eq:obj}. The remaining $3n$ components arise from the alignment penalty, where for each cable $i$:
\begin{equation}
\frac{\partial J}{\partial \alpha_i} = 2\mu\sum_{j \neq i}(\alpha_i^T \alpha_j)\alpha_j
\label{eq:alpha_gradient}
\end{equation}
The coupling between the cable directions through the alignment penalty in \eqref{eq:obj} enables collision avoidance behavior. Each unit norm constraint \eqref{eq:unit_norm} involves only its corresponding direction vector, giving the problem a block-diagonal Jacobian structure. For four UAVs this leads to 16 decision variables (four tensions and twelve direction components), which is small enough for real-time implementation. More generally, the number of variables increases linearly with the size of the team, preserving the same structural properties that allow efficient active set strategies to handle the tension bounds \eqref{eq:tension_bounds}.

\section{Results}\label{sec:results}
\begin{figure}[t]
 \centering
 \begin{subfigure}[b]{0.49\textwidth}
  \includegraphics[width=0.99\textwidth]{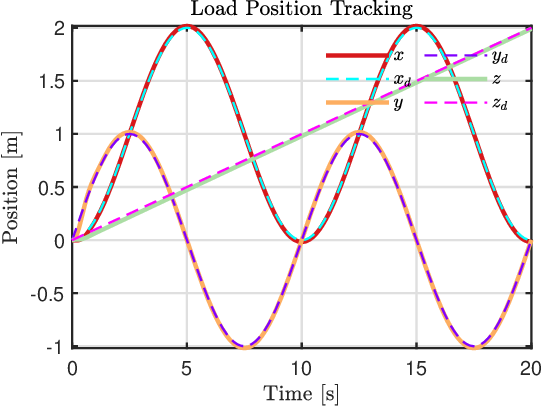}
  \caption{Trajectory tracking: reference vs actual}
 \end{subfigure}
 \hfill
 \begin{subfigure}[b]{0.49\textwidth}
  \includegraphics[width=0.99\textwidth]{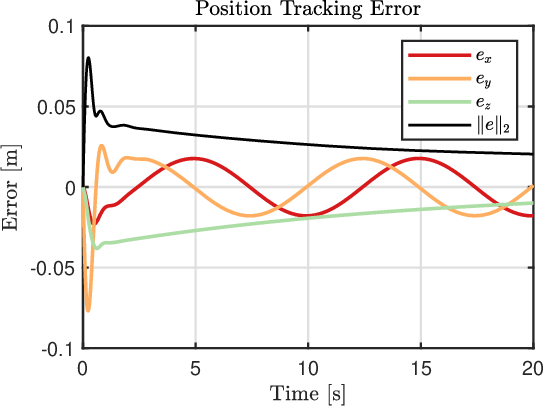}
  \caption{Position error components over time}
 \end{subfigure}
 \caption{Payload trajectory tracking performance during the ascending spiral mission. The SQP-based controller maintains tight tracking with RMS error of \SI{2.97}{\centi\meter}.}
 \label{fig:traj_comparison}
\end{figure}
All simulations were carried out in MATLAB/Simulink using a fixed-step Runge– Kutta solver with a \SI{1}{\milli\second} integration step. The system consisted of four quadrotors ($m_i = \SI{0.5}{\kilogram}$, $J = \mathrm{diag}(2.1,\,1.87,\,3.97)\times 10^{-2}\,\si{\kilogram\meter\squared}$) transporting a \SI{0.225}{\kilogram} payload via \SI{1.0}{\meter} cables. Load controller gains: $k_{pL}=8.0$, $k_{dL}=2.0$; UAV controllers: $k_{pi}=40$, $k_{di}=10$, $k_{ii}=2$. These values were initialized from~\cite{oliva2024} for consistency, then refined in simulation for stable tracking.
\begin{figure}[t]
 \centering
 \begin{subfigure}[b]{0.49\textwidth}
  \includegraphics[width=0.99\textwidth]{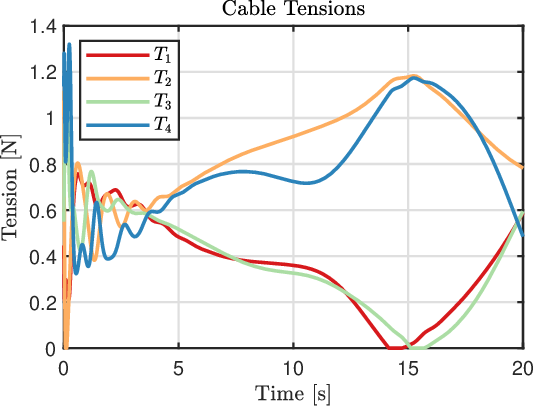}
  \caption{Without optimization}
 \end{subfigure}
 \hfill
 \begin{subfigure}[b]{0.49\textwidth}
  \includegraphics[width=0.99\textwidth]{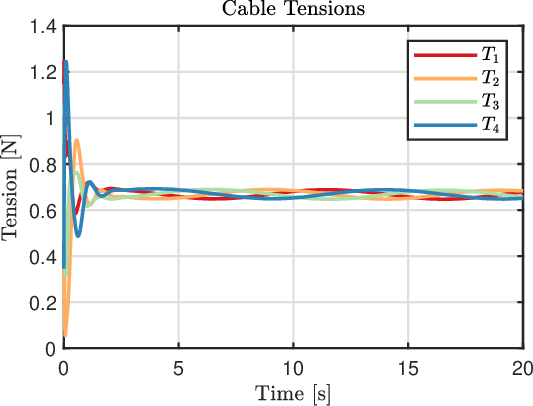}
  \caption{With SQP optimization}
 \end{subfigure}
 \caption{Cable tension profiles comparison. The SQP allocator (b) achieves significantly more balanced force distribution compared to the baseline geometric pattern approach (a).}
 \label{fig:tension_comparison}
\end{figure}

The reference trajectory is an ascending spiral in which the payload describes a circular path in the horizontal plane while gradually increasing altitude. Figure~\ref{fig:3d_config} shows a snapshot of the four-UAV system during this mission, where the optimization layer maintains a safe spatial configuration around the payload.
\begin{figure}[t]
 \centering
 \begin{subfigure}[b]{0.49\textwidth}
  \includegraphics[width=0.99\textwidth]{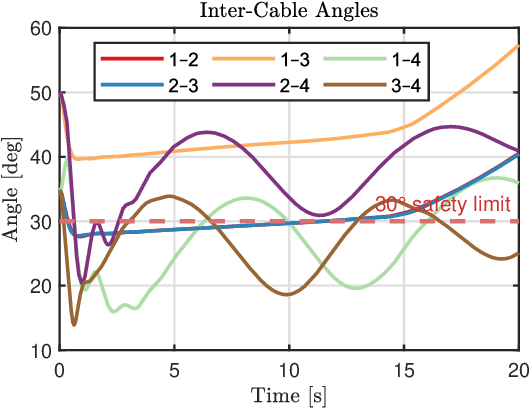}
  \caption{Without optimization.}
 \end{subfigure}
 \hfill
 \begin{subfigure}[b]{0.49\textwidth}
  \includegraphics[width=0.99\textwidth]{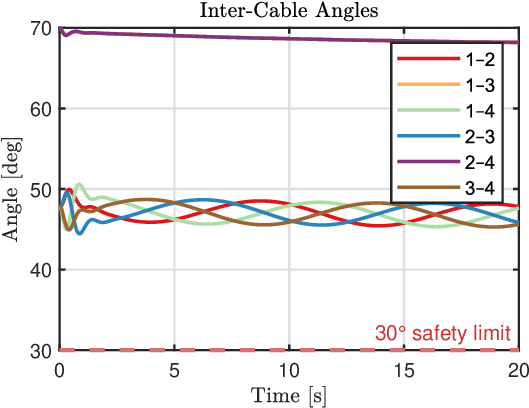}
  \caption{With SQP optimization.}
 \end{subfigure}
 \caption{Pairwise cable angles comparison. The SQP allocator (b) successfully prevents near-parallel cable configurations, maintaining all angles above \SI{44}{\degree}, while the baseline approach (a) frequently exhibits angles below the \SI{30}{\degree} safety threshold (shown as dashed line).}
 \label{fig:angles_comparison}
\end{figure}

Trajectory tracking performance demonstrates that centimeter-level accuracy is preserved with the SQP allocator. Figure~\ref{fig:traj_comparison} presents the reference and actual trajectories along with the corresponding position errors. The load remains tightly on course throughout the \SI{20}{\second} mission, with a root mean square (RMS) position error of \SI{2.97}{\centi\meter} and a maximum excursion never exceeding \SI{8}{\centi\meter}. This tracking accuracy matches that reported in previous MAATS studies that relied on simpler, non-optimized force allocators \cite{Jackson2020,oliva2024}, confirming that the optimization layer does not compromise control performance.
\begin{figure}[t]
 \centering
 \begin{subfigure}[b]{0.49\textwidth}
\includegraphics[width=0.99\textwidth]{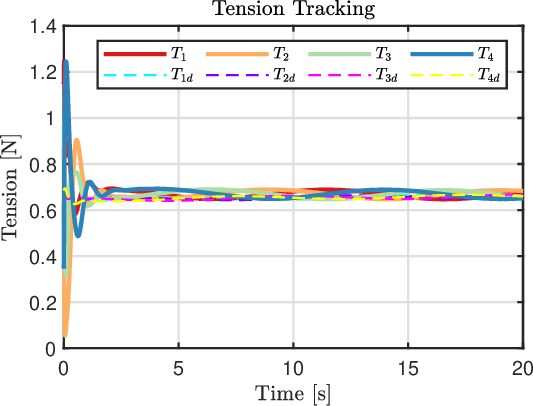}
  \caption{Desired vs actual tensions}
 \end{subfigure}
 \hfill
 \begin{subfigure}[b]{0.49\textwidth}
  \includegraphics[width=0.99\textwidth]{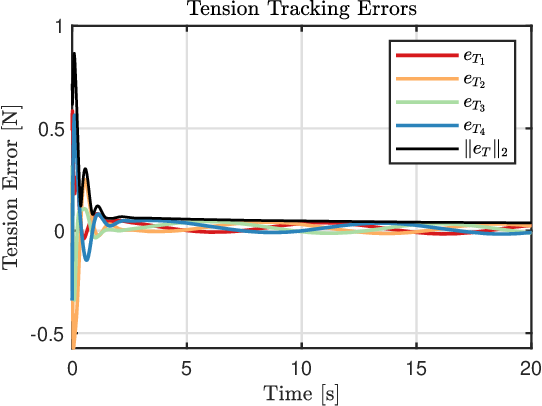}
  \caption{Tension tracking errors for each cable}
 \end{subfigure}
 \caption{Tension tracking performance showing the agreement between SQP-computed desired tensions and actual tensions achieved by the controllers.}
 \label{fig:tension_tracking}
\end{figure}

The effectiveness of the optimization approach becomes evident when comparing tension profiles with and without the SQP allocator. Figure~\ref{fig:tension_comparison} illustrates how the baseline controller \cite{oliva2024} produces an uneven tension distribution, with one cable carrying significantly more load than others. In contrast, the SQP-based allocation achieves a balanced force distribution among all four UAVs, with tension fluctuations restricted to a small range. This balanced distribution ensures more uniform battery discharge throughout the fleet and reduces the risk of individual motor saturation. The alignment penalty delivers a clear safety benefit, as demonstrated in Figure~\ref{fig:angles_comparison}. The SQP allocator with alignment weight \(\mu = 0.15\) maintains all pairwise cable angles comfortably above the \SI{30}{\degree} safety threshold throughout operation. The worst-case simultaneous total tension peaks at \SI{3.33}{\newton}, 1.24 times its long-term average, while the minimum angle between two cables never drops below \SI{44.4}{\degree}. The controller's ability to track desired tension commands is illustrated in Figure~\ref{fig:tension_tracking}, which compares the desired tensions computed by the SQP allocator with the actual tensions achieved by the controllers. The close alignment between the desired and actual values confirms that the hierarchical control architecture successfully implements the optimal allocation decisions. Minor discrepancies arise mainly from the transitory state, but these remain within acceptable bounds.
\begin{figure}[t]
 \centering
 \begin{subfigure}[b]{0.45\textwidth}
  \includegraphics[width=0.99\textwidth]{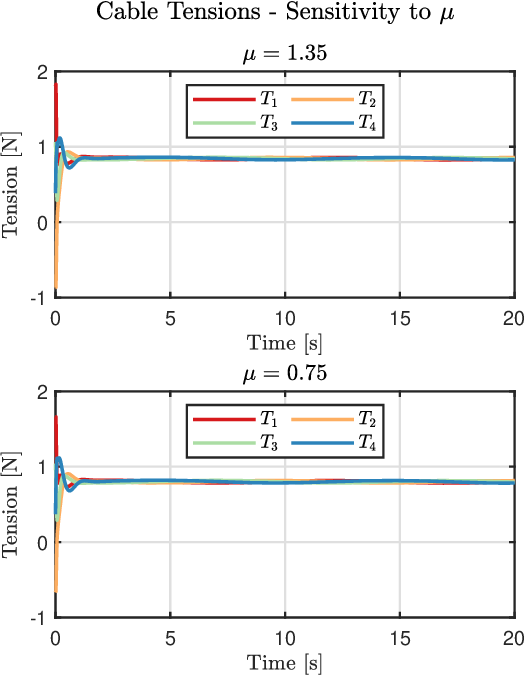}
  \caption{Cable tensions for different \(\mu\) values}
 \end{subfigure}
 \hfill
 \begin{subfigure}[b]{0.49\textwidth}
  \includegraphics[width=0.99\textwidth]{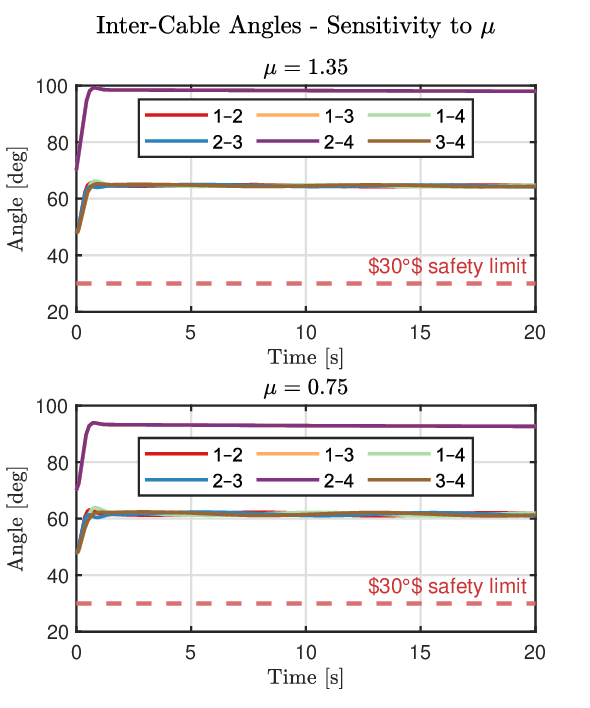}
  \caption{Cable angles for different \(\mu\) values}
 \end{subfigure}
 \caption{Sensitivity analysis showing the effect of alignment weight \(\mu\) on system performance. Higher \(\mu\) values increase cable separation at the expense of higher tensions, providing a tunable trade-off between safety and efficiency.}
 \label{fig:mu_sensitivity}
\end{figure}

The sensitivity of system performance to alignment weight $\mu$ provides a practical tuning knob for system safety. To gauge how much total cable force the agents expend over a mission, we use the integrated tension cost
$
  J_T \;=\; \int_{t_0}^{t_f}\sum_{i=1}^{n} |T_i(t)|\,dt,
$
a scalar that measures the cumulative magnitude of all cable forces throughout the mission interval $[t_0,t_f]$.

Figure~\ref{fig:mu_sensitivity}(a) compares the cable-tension profiles obtained with three alignment weights $\mu\!\in\!\{0.15,0.75,1.35\}$, while Fig.~\ref{fig:mu_sensitivity}(b) shows the corresponding inter-cable angles.  For these values, the minimum angles and integrated tension costs are $\min\theta =(\SI{44.4}{\degree},\,\SI{47.8}{\degree},\,\SI{47.9}{\degree})$ and $J_T =(26.8,\,32.1,\,33.7)\,\text{N·s}$, respectively.

Computational performance measurements confirm the real-time feasibility of the approach. Timing tests on a Windows 11 laptop with an Intel Core Ultra 7-155U processor show that a complete SQP cycle averages \SI{1.13}{\milli\second}, with 99th percentile and worst-case times of \SI{3.06}{\milli\second} and \SI{12.29}{\milli\second}, respectively. These metrics leave comfortable margins for communication layers on the same processor, making the approach viable for embedded implementation.

A key advantage of the SQP formulation is that it does not prescribe a fixed geometric pattern: starting from any feasible initial state, the optimizer converges to a safe and balanced configuration as long as force balance is satisfied. This contrasts with formation-based methods, which may fail when obstacles or operational constraints prevent maintaining a predefined shape.

The present study is limited by the assumption of rigid and massless cables and by the use of a centralized optimization layer, which introduces a single point of failure. In addition, results are based solely on simulations; hardware validation is required to assess performance under wind disturbances, delays, and modeling errors.

\section{Conclusion}\label{sec:conclusion}

This work introduced a lightweight SQP allocator that distributes cable tensions while penalizing small inter-cable angles. Integrated into a four-layer MAATS controller, the method maintains the payload trajectory, balances UAV effort, and respects real-time constraints on standard hardware. Because no formation is imposed, the controller adapts to different initial layouts, letting the optimizer find a safe, energy-efficient arrangement. Demonstrations with four UAVs confirmed feasibility, while the formulation naturally scales with team size, as each additional UAV introduces one tension and three direction variables. A sensitivity analysis further showed that the alignment weight provides a tunable trade-off between safety and energy with only marginal impact on tracking.

Future work will include Software-In-The-Loop and hardware experiments, as well as expanding the evaluation to larger UAV teams to assess scalability. Extending the payload model to a rigid body with full attitude dynamics would enable more complex cooperative transport tasks, while distributed optimization strategies could improve resilience by eliminating the single point of failure inherent in centralized allocation.

\subsubsection*{Disclosure of Interests.} The authors have no competing interests to declare that are relevant to the content of this article.

\bibliographystyle{splncs04}
\bibliography{suspendedLoad}

\begin{thebibliography}{10}
\providecommand{\url}[1]{\texttt{#1}}
\providecommand{\urlprefix}{URL }
\providecommand{\doi}[1]{https://doi.org/#1}

\bibitem{Cai2024}
Cai, J., Xian, B.: Robust hierarchical geometry control for a multiple‐uav aerial transportation system with a suspended payload. Nonlinear Dynamics  \textbf{112}(6),  4551--4571 (2024)

\bibitem{Doakhan2023}
Doakhan, M., Kabganian, M., Azimi, A.: Robust adaptive control for formation-based cooperative transportation of a payload by multi-quadrotors. European Journal of Control  \textbf{69},  100763 (2023)

\bibitem{Foehn2016}
Foehn, P., Falanga, D., Kuppuswamy, N., Tedrake, R., Scaramuzza, D.: Fast trajectory optimization for agile quadrotor maneuvers with a cable-suspended payload. In: Robotics: Science and Systems (RSS) Workshop (2016), workshop contribution

\bibitem{Geng2020}
Geng, J., Langelaan, J.W.: Cooperative transport of a slung load using load-leading control. Journal of Guidance, Control, and Dynamics  \textbf{43}(7),  1313--1331 (2020)

\bibitem{Goodarzi2015}
Goodarzi, F.A., Lee, T.: Dynamics and control of quadrotor uavs transporting a rigid body connected via flexible cables. In: 2015 American Control Conference (ACC). pp. 4677--4682 (2015)

\bibitem{Jackson2020}
Jackson, B.E., Howell, T.A., Shah, K., Schwager, M., Manchester, Z.: Scalable cooperative transport of cable-suspended loads with {UAVs} using distributed trajectory optimization. IEEE Robotics and Automation Letters  \textbf{5}(2),  3368--3374 (2020)

\bibitem{Li2023}
Li, G., Liu, X., Loianno, G.: Rotortm: A flexible simulator for aerial transportation and manipulation. IEEE Transactions on Robotics pp. 1--20 (2023)

\bibitem{MatlabSQP}
{MathWorks}: Constrained nonlinear optimization algorithms (2025), \url{https://www.mathworks.com/help/optim/ug/constrained-nonlinear-optimization-algorithms.html}, accessed: 12 Jun 2025

\bibitem{Maza2010}
Maza, I., Kondak, K., Bernard, M., Ollero, A.: Multi-{UAV} cooperation and control for load transportation and deployment. Journal of Intelligent \& Robotic Systems  \textbf{57}(1--4),  417--449 (2010)

\bibitem{Mercado2024}
Mercado-Ravell, D., Oliva-Palomo, F., Sanahuja, G., Castillo, P.: Control and real-time experiments for a multi-agent aerial transportation system. Journal of the Brazilian Society of Mechanical Sciences and Engineering  \textbf{46}(12), ~616 (2024)

\bibitem{oliva2024}
Oliva-Palomo, F., Mercado-Ravell, D., Castillo, P.: Aerial transportation control of suspended payloads with multiple agents. Journal of the Franklin Institute  \textbf{361}(7),  106787 (2024)

\bibitem{Tognon2018}
Tognon, M., Gabellieri, C., Pallottino, L., Franchi, A.: Aerial co-manipulation with cables: The role of internal force for equilibria, stability, and passivity. IEEE Robotics and Automation Letters  \textbf{3}(3),  2577--2583 (2018)

\bibitem{VillaBrandao2020}
Villa, D.K.D., Brandão, A.S., Sarcinelli-Filho, M.: A survey on load transportation using multirotor {UAVs}. Journal of Intelligent \& Robotic Systems  \textbf{98}(2),  267--296 (2020)

\bibitem{Villa2021}
Villa, D.K.D., Brandão, A.S., Carelli, R., Sarcinelli-Filho, M.: Cooperative load transportation with two quadrotors using adaptive control. IEEE Access  \textbf{9},  129148--129160 (2021)

\bibitem{Wahba2024}
Wahba, K., Hönig, W.: Efficient optimization-based cable force allocation for geometric control of a multirotor team transporting a payload. IEEE Robotics and Automation Letters  \textbf{9}(4),  3688--3695 (2024)

\bibitem{Wehbeh2020}
Wehbeh, J., Rahman, S., Sharf, I.: Distributed model predictive control for {UAVs} collaborative payload transport. In: 2020 IEEE/RSJ International Conference on Intelligent Robots and Systems (IROS). pp. 11666--11672 (2020)

\end{thebibliography}

\end{document}